\documentclass[
    ,final            % use final for the camera ready runs
  ]
  {aipproc}

\layoutstyle{8x11double}

\usepackage{amsmath}
\usepackage{graphicx}% Include figure files
\usepackage{dcolumn}% Align table columns on decimal point
\usepackage{bm}% bold math
\usepackage{enumerate}
\usepackage{ctable}
\usepackage{booktabs}
\PassOptionsToPackage{super,compress}{natbib}
\usepackage{siunitx}

\newcommand{\JM}{\textit{Just Math}}

\begin{document}
\title{Just Math: A New Epistemic Frame}

\author{Steven F.~Wolf}{
	address = {CREATE for STEM Institute, 
	Michigan State University, East Lansing, MI 48824, USA},
}%

\author{Leanne Doughty}{
	address = {Department of Physics and Astronomy,
	Michigan State University, East Lansing, MI 48824, USA},
}%

\author{Paul W.~Irving}{
	address = {Department of Physics and Astronomy,
	Michigan State University, East Lansing, MI 48824, USA},
	altaddress = {CREATE for STEM Institute, 
	Michigan State University, East Lansing, MI 48824, USA}
}%

\author{Eleanor C.~Sayre}{
	address = {Department of Physics, 
	Kansas State University, Manhattan, KS 66506, USA},
}%

\author{Marcos D.~Caballero}{
	address = {Department of Physics and Astronomy,
	Michigan State University, East Lansing, MI 48824, USA},
	altaddress = {CREATE for STEM Institute, 
	Michigan State University, East Lansing, MI 48824, USA}
}%

\date{\today}

\begin{abstract}
A goal of PER is to understand how students use math in physics contexts.  To investigate how students use math, we need to identify transitions between conceptual sense-making about physical systems and using mathematics to describe and to make predictions about those systems. We reviewed video of students solving a variety of physics problems in small groups through the lens of epistemic frames (e-frames).   In this paper, we present a new e-frame, which we are calling ``Just Math'', that is similar to the Worksheet e-frame, but is characterized by brief, low-level, math-focused utterances between students, in some cases along with expansive off-topic discussions. Future work will focus on analyzing the transitions into and out of this e-frame so that we may develop a more coherent understanding of students' use of math in physics.
\end{abstract}

\classification{01.40.Fk, 01.40.Ha}
\keywords{Research in physics education, Learning theory and science teaching}

\maketitle

\section{Introduction}
Physics provides a rich environment to investigate how students use mathematics in science and engineering contexts. Many of the models developed in physics are quantitative and require students to use sophisticated mathematics (e.g., vector calculus) to develop a deep understanding of those models. Furthermore, physics courses are often the gateway to using math in the way that science and engineering professionals do. It is this skillful use of mathematics within physics courses that is a large portion of the challenge for STEM majors.

Research on student use of mathematics in physics frequently focuses on specific physics or mathematical tools (e.g., \cite{Zhu:2012vo,PhysRevSTPER.8.023101}).  By contrast, other work seeks to develop more general constructs for understanding this vital aspect of students' development (e.g., \cite{Wilcox:2013ea,Hammer:2000vf,PhysRevSTPER.4.020105,PhysRevSTPER.9.020108}), often focusing on describing students' in-the-moment reasoning, not the mathematical tools themselves.  
%Our work brokers the combination of these approaches (making them more consonant with each other) in order to develop a coherent, experimentally-validated, and predictive theory of students' use of mathematics in physics that can be used across contexts, modalities, and levels.

Our work grows out of this latter tradition. In the course of analyzing video of upper-division students solving problems {\it in situ}, we noticed that they shift between sense-making activities and activities whose sole purpose appears to be ``doing math.'' We are interested in investigating: how do students collectively set up the problem and determine that they are ready to "just do math"? how (if at all) do students reflect upon and interpret the result they get from doing the math?  To answer these questions we need to be able to reliably identify when students are ``just doing math." To this end, we have conducted a preliminary study where we leveraged the epistemic frame construct \cite{Scherr:2009gm,Irving:2013bi}. We have identified markers in students' behaviors and discourse that are indicative of students "just doing math."

In this paper, we describe the features of this new epistemic frame (e-frame), \JM. We distinguish \JM~from a related e-frame and discuss how this distinction has important implications for our future work.

% We are curious:  What are students' expectations and behaviors when they ``just do math''? What prompts students to switch into and out of these ``just math'' behaviors?

%In this paper, we describe an epistemic frame that is endemic to students' problem-solving work: \JM.   

%In this paper, we describe the first steps we have taken to develop our theory. We have observed groups of students working on physics problems {\it in situ}. Through these observations, we discovered students engaging in what appears to be a universal and inevitable activity: they do math. Watching students execute mathematical procedures is certainly not interesting on its own, but reliably identifying when students are doing math allows us to determine what tips students into and out of doing math. 

%In this paper, we use the lens of epistemic framing \cite{Scherr:2009gm,Irving:2013bi} to introduce a new epistemic frame (e-frame), \JM, to account for students' behaviors and expectations about doing math in physics contexts.  This preliminary study focuses on the features of this new e-frame and how we distinguish it from a related e-frame. This work is grounded in observations of upper-division students working on homework problems together in groups.

\section{Data Source and Context}
Our data comes from naturalistic observations of homework help sessions (HHS) in an upper-division electricity and magnetism course. The HHS were an informal weekly meeting where students could work with their colleagues to solve homework problems. The HHS were facilitated by two teaching assistants (TAs) who guided discussion but did not present problems or their solutions. Typically between 5 and 10 students (split into two cooperative groups) attended the session; a core group of 5 students attended regularly. Two pairs of two unattended video cameras recorded each group. The first camera recorded from a distance to capture gestures while the second camera recorded from a high angle above the table to capture students writing on a table-based whiteboard. In total 114 hours of video was captured. The data presented in this paper is focused on 9 hours of HHS data encompassing 3 different groups of students.

%In this work, we have employed the lenses of epistemic frames \cite{Scherr:2009gm} and framing \cite{Irving:2013bi} to allow us to use both the behaviors of students and their talk to identify the link between how the actions of students inform their expectations for a group activity.  With these theoretical tools in mind, we have observed video of students working on different mathematical tasks in order to understand how students are using math in physics.

%Using these theoretical tools has helped us identify when students are just doing math, which appears to be a universal and inevitable activity.  Moreover, we propose that doing math is an e-frame, which we are naming \JM.  Watching students execute mathematical procedures is certainly not interesting on its own.  However, reliably identifying \JM~allows us to determine what tips students into and out of \JM.  After all, our students' skill in using math in physics contexts will depend on how we tip students into using math in different and productive ways.

\section{Theoretical Framework}
Framing is the combination of several resources in an effort to understand and work within a situation \cite{PhysRevSTPER.4.020105}. Epistemic framing differs as it is a perception (unconscious or conscious) of the tools and skills required in a particular context or situation. E-frames can be envisioned as a storage area for conceptual and procedural resources, promoting some resource's activation and blocking others \cite{Irving:2013bi}. We have drawn from the literature on e-frames because we believe that when students are doing \JM, they perceive the activity as requiring a particular set of behaviors and math oriented resources. 

Building on the work of \citeauthor{Scherr:2009gm}, we studied the behavioral patterns that students exhibit when working in group settings. In an introductory tutorial setting, \citeauthor{Scherr:2009gm} identified four e-frames (the \textit{Discussion} e-frame, the \textit{TA} e-frame, the \textit{Joking} e-frame and the \textit{Worksheet} e-frame) that they characterized by coding for specific behaviors and expectations \cite{Scherr:2009gm}. For example, in the \textit{Worksheet} e-frame, students focused their gaze on their personal worksheet, and they hunched their bodies over their paper. Student talk came in short bursts and was monotone.  \citeauthor{Scherr:2009gm} argue that these behaviors were indicative that each member of the group expected to work on a personal task, specifically, answering the question on the worksheet.

Distinguishing \JM~from other e-frames (especially the \textit{Worksheet} e-frame) requires an analysis of student discourse.  \citeauthor{Irving:2013bi} analyzed the content of student talk along two axes \cite{Irving:2013bi}:  ``narrow vs.~expansive'' and ``silly vs.~serious''.  ``Narrow vs.~expansive'' focuses on the scope of student discussion, while ``silly vs.~serious'' focuses more on how playfully the students interact while in a \textit{Discussion} e-frame. \citeauthor{Irving:2013bi} found that when students work towards the goal of solving a problem, their talk is most often both narrow and serious. %\mdcsays{We need to deal with the use of the word ``framing'' here.}  \swsays{I changed framing to dimension and re-worked the first sentence so that we didn't overuse the word ``dimension.''}

\section{Analysis}

By combining e-frames with two-axis framing, we identified the \JM~frame. From the perspective of student behavior, \JM~is similar to the \textit{Worksheet} e-frame: a hunched posture, hands busy writing, gaze focused on the page, and discourse that is typically brief. However, applying the two framing axes to the content of students' talk distinguishes \JM~from the \textit{Worksheet} e-frame. We coded 9 hours of video for student behaviors, discourse, and expectations. 

The coding process began by first focusing on coding the HHS data for student behaviors. Two researchers (SFW and LD) independently reviewed approximately 9 hours of HHS data looking for postures, gestures, and gaze patterns consistent with the \textit{Worksheet} and \JM~e-frames. These researchers negotiated (with each other and, later, with members of the team) the start and stop times associated with what they believed were candidate episodes of \JM. After negotiation, we found approximately 1.5 hours of candidate HHS data where students exhibited behaviors consistent with the \textit{Worksheet} or \JM~e-frames.

Having identified candidate episodes using an analysis of student behaviors, we analyzed the content of student talk during these candidate episodes. Episodes where student talk included narrow and serious {\it math-focused} check-ins (see Episode 1) and/or expansive and silly comments (see Episode 2) were identified as \JM. These episodes were discussed and the relevant features negotiated among members of the team. 

If there was insufficient evidence of an episode being \JM~from the analysis of the discourse and behaviors then the analysis process switched to expectations, where the coders micro-analyzed the behaviors and discourse  before and after candidate \JM~episodes for the groups expectations.  The combined analysis for behaviors, discourse and expectations resulted into the identification of approximately 10 minutes of HHS data that could be definitively categorized as \JM. We believe that, to date, there are unidenitfied episodes of \JM~present within the 1.5 hours of candidate HHS data but there are a number of  challenges to definitively identifying episodes of \JM, which will be discussed later. In the following sections, we discuss the three elements of the \JM~e-frame through examples drawn from our data. %\swsays{We changed the subtitles from ``behaviors, discourse, and expectations'' to ``behaviors, DISCOURSE, and expectations'', but didn't change the all of the places we put that phrase in this section.  I've fixed that here.}

%Due to the nature of the worksheet e-frame and the \JM~ e-frame it can be hard to interpret how students are framing an activity. The etymology of the worksheet e-frame can be traced back to the activity the students are engaged in during the period being coded: ``filling in the worksheet." The physical artifact of the worksheet is where students were focusing their attention, and the lack of verbal exchanges during the frame would have made it difficult to analyze the period in a more meaningful way.

\subsection{Behaviors} 
\label{ssec:beh}

\begin{figure*}
	\includegraphics[width=\textwidth]{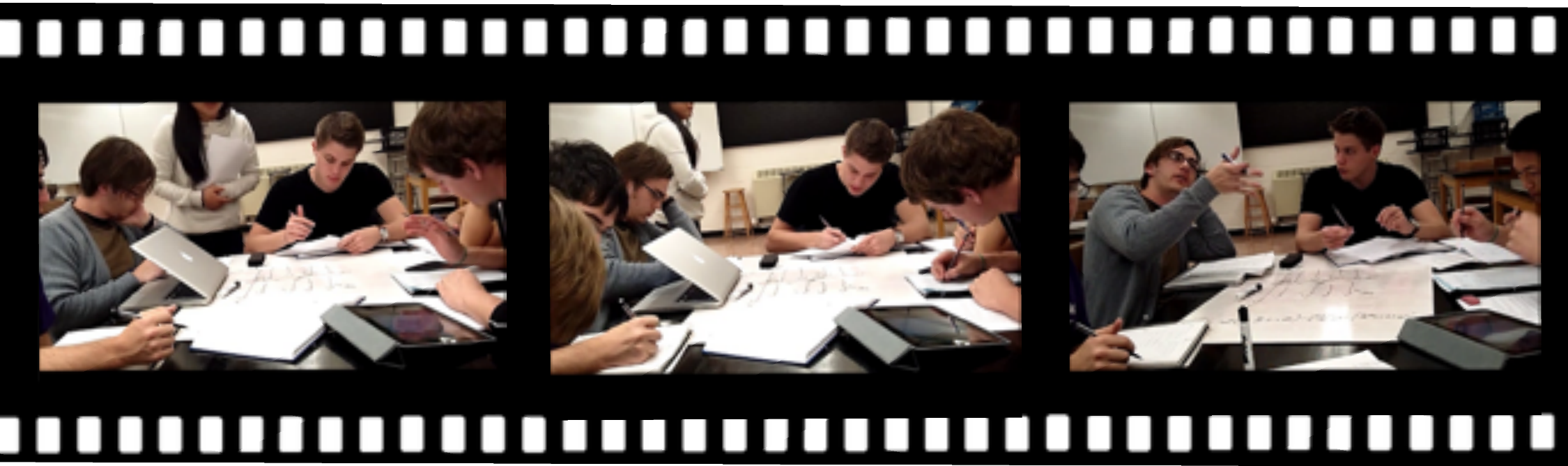}
	\caption{Group body language before and during \JM~epistemological frame - Jonah group week 3. In the left frame, students focus is shifting to their own papers after engaging in a group discussion.  In the center, we see students exhibiting \JM~behaviors.  On the right, students have exited \JM, and are now discussing their solutions to the problem.}% \swsays{Need higher-res figure.}}
	\label{fig:group}
\end{figure*}

The behaviors observed while students are in the \JM~e-frame are similar in nature to the behaviors described for the previously discussed \textit{Worksheet} e-frame. Figure \ref{fig:group} illustrates the posture of the students before, during, and after \JM.  In the left frame, students are just beginning to enter \JM; they hunch over their respective papers and shift their gaze from the group and its resources to their own papers.  In the middle frame, we see the students' hands are busy writing.  Some students pause briefly to check in with the group or a personal resource, such as a textbook, and then return to writing.  In the right frame, we see that students have left \JM; they are now sitting upright and looking at each other while discussing their solutions to the problem. %\mdcsays{Label Figure 1's frames 1a, 1b, and 1c and use those references here?}

% \begin{figure}
% 	\includegraphics[width=\columnwidth]{mla_wpm.pdf}
% 	\caption{Words per minute for the magnetostatics episode.  From the fifth through the eighth minute, we identify the group's frame to be \JM~- Larry group week 5.}
% 	\label{fig:wpm1}
% \end{figure}

\subsection{Discourse} \label{ssec:disc}
In order to distinguish \JM~from the \textit{Worksheet} e-frame, we analyzed the students' discourse.  First, we noticed that the rate of talk decreases when students are engaged in \JM.  Furthermore, the nature of that talk fell into one of two categories: 

(1) Narrow and serious \textit{math-focused} check-ins about the problem (e.g., the integral of sine is positive or negative cosine), which are brief in nature. In the following episode, 5 students are sitting around a table, working on a problem to determine if a particular static magnetic field has no curl.  Students begin the episode with a discussion about the curl including consulting the front flyleaf of the textbook. Once they begin taking the curl, students focus on writing on their own papers (left frame of Figure \ref{fig:group}). After a couple of minutes of writing, a student decides to check-in with his group:
\begin{description} 
	\itshape
	\item[Episode 1]  
	\item[L:] So this is zero because the phi component is constant with respect to z, right?
	\item[J:] Yeah
\end{description}
% \begin{description}
% \itshape
% \item [Episode 1]
% \item[A:] [Is the] derivative of sine negative cosine?
% \item[TA:] Yes
% \item[Group:] No! (laughing) Positive!
% \end{description}
After this brief exchange the group returns to non-communicative, paper-focused behavior (middle frame of Figure \ref{fig:group}).

% Or this exchange? (Different group)

(2) Expansive and silly (off-topic) comments while the students are working.  This second kind of talk was not reported in \citeauthor{Scherr:2009gm}'s \textit{Worksheet} e-frame and does not fit in the \textit{Joking} e-frame from a behavioral perspective. In the following episode, 3 students are working on a problem that is requiring them to take the curl of a magnetic field in cylindrical coordinates. Once the group has entered \JM, they begin a discussion that is off-topic:
\begin{description}
\itshape
\item[Episode 2]
\item[L:] Boom, good job Jonah, you're ahead of me on this
\item[J:] Well, I started in the complete wrong spot so\ldots
\item[L:] Have you and Ian busted out that telescope yet?
\end{description}
%\essays{If you reformat all the transcript like this, you'll save a few inches.}
From this introductory question an off-topic conversation about telescopes continues. For the whole exchange, the students continue to focus on their own personal papers, and do not share gaze with each other or attend to their joint whiteboard. All of the students continue to write and process through the remaining math of the problem while concurrently talking about telescopes. Both of these types of talk illustrated are characteristic of the \JM~e-frame. 

\subsection{Expectations} \label{ssec:exp}
%\mdcsays{Read through this section and make sure it makes sense. I did some combining and editing.}
Group expectations can be inferred from the behaviors and discourse of the group before, during, and after \JM.  The HHS is an environment where the expectation is to obtain a solution to the homework problem. The focus of the discussions prior to entering \JM~is on building a mathematical procedure or representation from the shared resources of the group. This procedure or representation is typically negotiated until the group reaches a point in time where each member is comfortable to proceed with ``doing the math.'' The following episode contains the discussion that occurs prior to the students entering \JM~in Episode 2, which was discussed previously.
\begin{description}   
	\itshape  
	\item[Episode 3]
	\item[J:] It gives us an equation for B\ldots and I think it's just find like\ldots take the curl of it\ldots but then the next part says suppose that the field decreases as $r^2$ instead of $r$, show that the curl is non-zero.
	\item[L:] Yeah but I mean if you look at the\ldots curl for it in cylindrical coordinates and you can show that\ldots the terms go to zero.
\end{description}
This discussion of the curl and how to take it in cylindrical coordinates continues for a few minutes until the participants are ready to do the math and enter into \JM. The scaffolding by the students through narrow/serious discussion to a point in the problem where they as a group can enter into \JM~is an essential element of our ability to identify the \JM~e-frame and distinguish it from the \textit{Worksheet} e-frame.  Once the rest of the group agrees, \JM~begins. Once the transition into \JM~has occurred, there are several expectations that become apparent and observable through the student's behavior and discourse. The main expectation being that to proceed with the problem they need to ``do math''. Students also have the expectation that ``doing math'' is a solitary activity. These expectations are illustrated by the math based check-ins (Episode 1) and the behavioral cues previously discussed. There is also an expectation that being in the \JM~e-frame is not as taxing as the discussion that preceded it and that they can partake in silly off-topic talk while processing through the math (Episode 2). 
There is also an expectation that it is possible that the scaffolding they have constructed is not mathematically sound and that bids to transition back to a \textit{Discussion} e-frame are allowed if a complication arises with the math.

\section{Discussion and Conclusions}
In this paper, we identified the \JM~e-frame, defined by students' behavior, discourse, and inferred expectations. The \JM~e-frame is strongly related to the \textit{Worksheet} e-frame \cite{Scherr:2009gm}, but also includes important differences. The behaviors that students engage in are typical of the \textit{Worksheet} e-frame. Distinctions between the \JM~frame and the \textit{Worksheet} e-frame are evident based on the content of the group discourse. This includes the possibility of off-topic conversation while processing through the math but more importantly the emphasis on math-oriented discourse to infer shared framing and expectations. Both the check-ins during \JM~and the scaffolding that occurs before transitioning are math centered. When students enter the \JM~e-frame one expectation is that in order to proceed with or finish the problem they must do math. This is in contrast to the \textit{Worksheet} e-frame where students' expectation is that they now need to write down an answer. This difference will affect what students do before they enter the frame and how they proceed when they exit the frame.  

%These differences imply that solving more canonical, mathematical physics problems is epistemically different from solving conceptual, tutorial-type problems. 

Finding the \JM~frame in natural observations presents some methodological challenges. Our analysis is necessarily confined to contexts where students work together in groups and talk about their problem solving.  While we believe that individuals may enter \JM~when working alone, our analysis has not uncovered that in a naturalistic setting.  Furthermore, even in groups, we are dependent on students' discourse to distinguish it from the \textit{Worksheet} e-frame. The check-ins that students make during \JM~are an important aspect of our ability to distinguish between it and the \textit{Worksheet} e-frame due to the overlapping behaviors between the two. However, check-ins only occur if the students encounter a problem. Encountering a problem does not always occur because the group may have scaffolded the \JM~frame adequately or may not need to check on a more trivial element of the math needed to solve the problem.  
%It is for this reason that the instances we definitively identified as \JM~tended to last for at least a minute, rather than a few seconds.  
Hence, the amount of data that we identified as \JM~is likely the lower measure for this data set. This point merely highlights the need to look at the e-frames from which students move into \JM~so that we can better identify when students are doing math. In the future we will characterize the pre and post \JM~behaviors, discourse, and expectations in greater detail in order to build up a library of before, during, and after indicators of \JM.

The \JM~e-frame is, by itself, of limited importance but it is the future work that its identification facilitates that makes its identification valuable. Transitions into and out of \JM~are a fruitful area of future research. Students' sense-making and scaffolding to bring their group to the point where it can enter \JM~is especially interesting, because research on how students use mathematics in group settings is sparse. Different amounts of scaffolding may need to be provided by and for different group members in order to result in a common negotiated \JM-ready group state. Identifying \JM~allows us to examine this negotiation and scaffolding process and the individual transitions into and out of \JM. The study of this combination of math related elements should provide greater insight into how students use math in physics contexts. \\

%\swsays{This gets back to $<4$ pages when we trim all of our comments.}

The authors would like to thank Ying Chen, who identified one of the examples of \JM~presented here and staffed the HHS for the course we observed. We also thank the members of PERL@MSU for their useful comments and suggestions on a draft of the manuscript.

\bibliographystyle{aipproc}
\bibliography{JustMathPERC}

\end{document}